\documentclass[aps,prl,groupedaddress,a4paper,preprint,endfloats]{revtex4}
\usepackage{bm,graphicx,graphics,amsmath,amssymb,bm,epsfig,color,pifont}
\usepackage{euscript,tabularx,textcomp}

\begin{document}
\bibliographystyle{apsrev}
 
\title{Magnetic resonance studies of the fundamental spin-wave modes
  in individual submicron Cu/NiFe/Cu perpendicularly magnetized
  disks.}

\author{G. de Loubens} \affiliation{Service de Physique de l'{\'E}tat
  Condens{\'e}, CEA Orme des Merisiers, F-91191 Gif-Sur-Yvette}
\author{V. V. Naletov} \thanks{Also at Physics Department, Kazan State
  University, Kazan 420008 Russia} \affiliation{Service de Physique de
  l'{\'E}tat Condens{\'e}, CEA Orme des Merisiers, F-91191 Gif-Sur-Yvette}
\author{O. Klein} \affiliation{Service
  de Physique de l'{\'E}tat Condens{\'e}, CEA Orme des Merisiers, F-91191
  Gif-Sur-Yvette}
\author{J. Ben Youssef} \affiliation{Laboratoire de Magn\'etisme de
  Bretagne, 6 Av. Le Gorgeu, F-29285 Brest}
\author{F. Boust} \affiliation{ONERA, 29 avenue de la Division
  Leclerc, F-92322 Ch\^atillon}
\author{N. Vukadinovic} \affiliation{Dassault Aviation, DGT/DTIAE, 78
  quai Marcel Dassault, F-92552 Saint-Cloud}

\date{\today}

\begin{abstract}
  Spin-wave spectra of perpendicularly magnetized disks consisting of a 100~nm permalloy (Py) layer sandwiched between two Cu layers of 30~nm are measured individually by a Magnetic Resonance Force Microscope (MRFM). Using 3D micromagnetic simulations, it is demonstrated that, for sub-micron size diameters, the lowest energy spin-wave mode of the saturated state is not spatially uniform, but rather is localized at the center of the Py/Cu interface in the region of the minimum demagnetizing field.
\end{abstract}


\maketitle

The detailed understanding of the spectrum and spatial structure of
spin-wave eigenmodes in submicron-size patterned heterostructures
\cite{hiebert97,bailleul06} introduces new challenges and
opportunities for novel magnetoelectronics devices \cite{wolf01}. The
exact nature of the lowest energy modes are of a particular interest
as they are the most susceptible to spin transfer excitation. In the
case of confined geometries, their identification can be quite
challenging because of the interplay between short-range exchange and
long-range dipolar interactions. Recent simulations predict that the
lowest energy mode of nanomagnets in a zero applied field is a mode
localized at the edges \cite{mcmichael05} instead of the uniform mode,
that has the lowest energy in larger structures. Furthermore
fundamental mechanisms specific to metallic multilayers (such has spin
accumulation at the boundary of normal and ferromagnetic metals
\cite{hurdequinticm}) can play a significant role in the magnetization
dynamics of multilayer metallic elements.

\begin{figure}
\includegraphics[width=10.0cm]{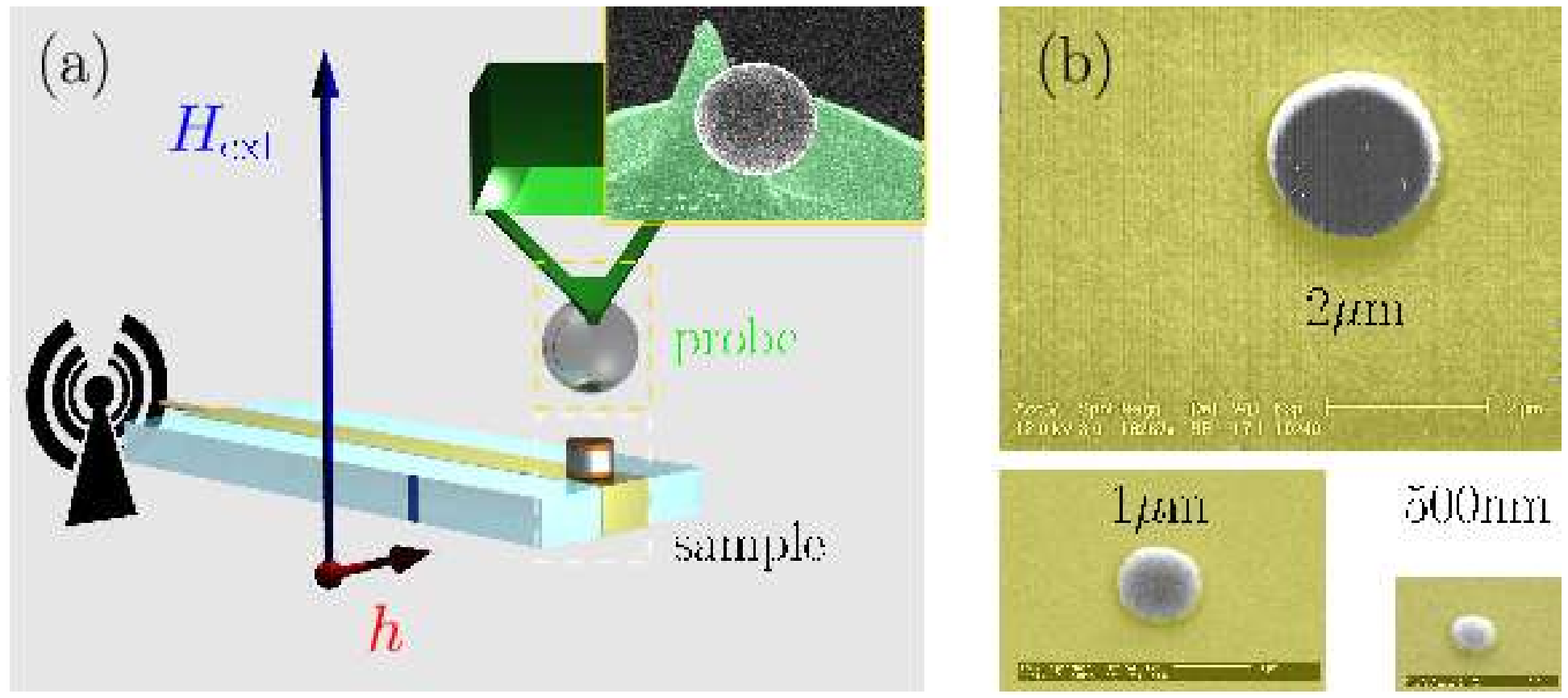}
\caption{ (a) Schematic of the setup showing SEM images of the
  mechanical probe and (b) of the disks.}
\label{fig1}
\end{figure}

Ferromagnetic resonance (FMR) is regarded as the basic tool to study
the microwave susceptibility of magnetic samples \cite{wigen84}. It
uses a well-defined excitation symmetry, where the microwave field $h$
couples to the most uniform mode in the sample.  However, the limited
sensitivity of standard FMR spectrometers implies that it can only be
performed on arrays of micron-size samples \cite{kakazei04}, which
statistically averages the spectra of each element and makes it
insensitive to individual differences. This Letter reports on the FMR
spectroscopy of \emph{individual} submicron heterostructures. To
detect their dynamical response, we exploit the high sensitivity of
mechanical-FMR setup \cite{zhang96, charbois02}. As shown
schematically in Fig.\ref{fig1}(a), the static part of the sample
magnetization, $M_z$, is coupled through the dipolar interaction to a
magnetic sphere attached at the end of a soft cantilever (spring
constant $5$~mN/m). Exciting the sample at a fixed frequency, the
spectrum is obtained by measuring the cantilever motion as a function
of the perpendicular dc applied field, $H_{\text{ext}}$. The force on
the cantilever is proportional to $\Delta M_z$ (the variation of the
component along the precession axis associated with the resonance).
The sphere is a metallic alloy whose principle constituents are Co (80
wt\%) and Fe (10 wt\%). It has a diameter of $4.0~\mu$m and its
magnetic moment is $(5\pm0.5) 10^{-9}$~emu (saturated above $3.8$
~kOe).  The center of the sphere is positioned above the center of the
disk $(3.1\pm0.1)~\mu$m away from the sample surface. In this
position, the distortion induced by the stray field of the sphere on
the FMR spectrum corresponds to a shift of $300\pm20$~G of the whole
spectrum \cite{charbois02}. To enhance the sensitivity by the quality
factor of the cantilever ($Q=4500$), a lock-in is used to measure the
mechanical response to an amplitude modulated excitation, where the
modulation frequency is set at the resonance frequency of the
cantilever (well below all the relaxation rates in the spin system).
It should be noted that the mechanical signal represents the energy
\emph{stored} in the spin-wave system, which is equal to the
\emph{absorbed} power ($P_\text{abs} \propto \chi''$, the microwave
susceptibility) during the spin-lattice relaxation time.  Thus the
mechanical-FMR spectrum is composed of bell-shaped resonances, much
like an absorption spectrum ($ \chi''$ vs.  $H_{\text{ext}}$), except
that the amplitude of each peak is renormalized by their respective
relaxation rate \cite{klein03}.

The excitation antenna is a 50~$\Omega$ microstrip circuit shorted at its
extremity. It is a Ti/Au(150~nm) line deposited on a
sapphire substrate.  Once completed, the microwave circuit is placed
inside a sputtering chamber to deposit the metallic trilayer, which
consists of a single Permalloy (Py) magnetic layer, $t=100$~nm in
thickness, sandwiched by two normal metal layers made of 30~nm thick
Cu. Subsequently, several Cu/NiFe/Cu disks with diameters ranging from
2~$\mu$m down to $0.5~\mu$m (\textit{cf.}  Fig.\ref{fig1}(b)) are patterned
by electron-beam lithography and ion-milling techniques out of the
same thin film. All the disks are placed at a magnetic field anti-node
of the microstrip.  A 50~$\mu$m separation is set between them, so that
any inter-disk coupling can be neglected.

\begin{figure}
\includegraphics[width=10.0cm]{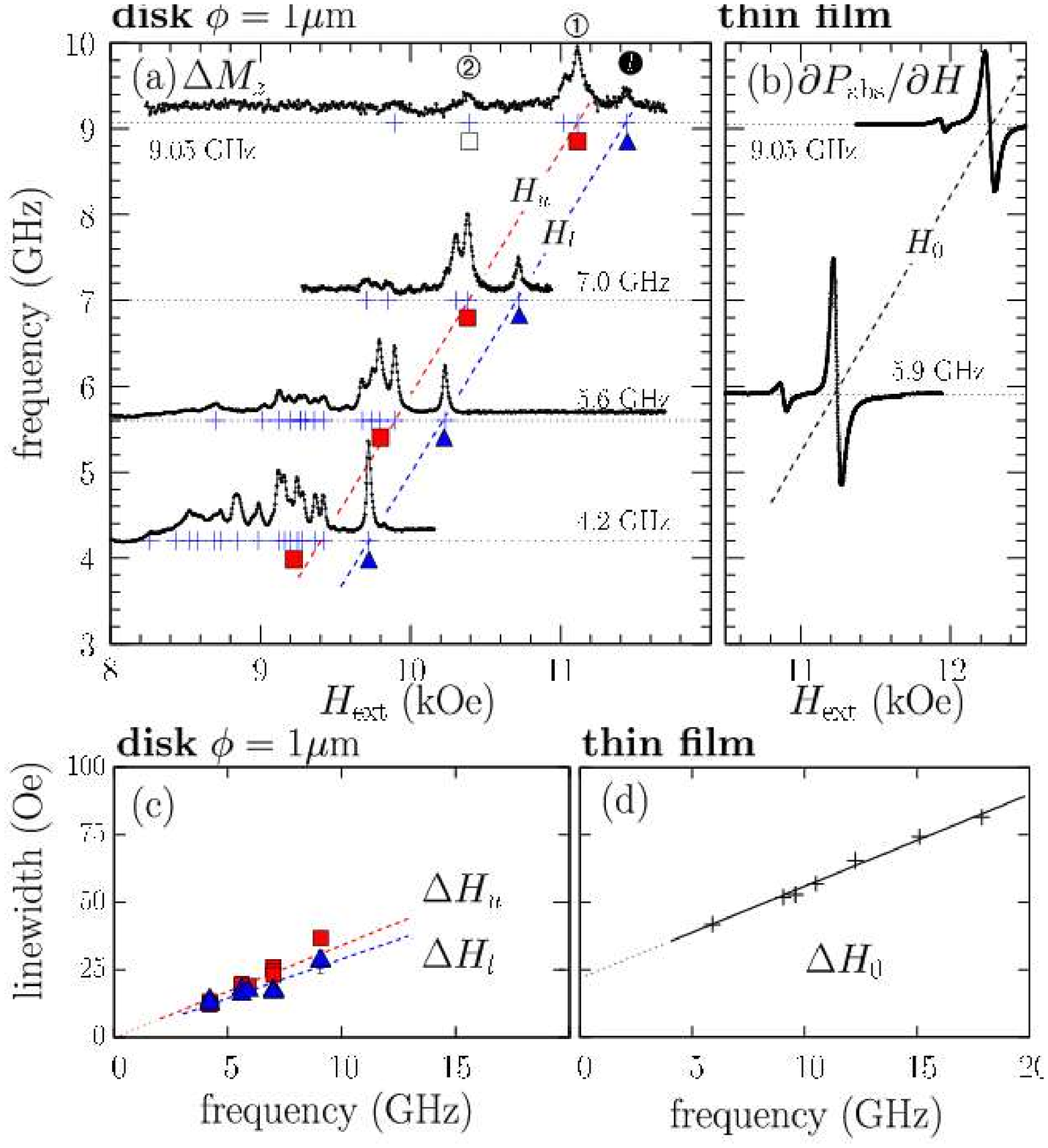}
\caption{(a) Mechanical-FMR spectra of the $\phi=1 ~\mu$m disk for
  different frequencies. The experimental positions of the two lowest
  energy modes are shown as squares and triangles. (b) Stripline-FMR
  spectra of the control thin film on Si at two frequencies. The
  bottom figures show the linewidth versus frequency for the two
  lowest energy modes of the disk (c) and extended film (d).}
\label{fig2}
\end{figure}

We first characterize the properties of a $\phi = 1 ~\mu$m diameter
disk.  For all the mechanical-FMR measurements presented herein: (a)
the data are collected at T=280~K in the linear regime, where the peak
amplitude remains proportional to the excitation power \textit{i.e.}
precession angles limited to 1\textdegree. (b) The disks are almost
saturated in the field range measured. (c) The shift introduced by the
stray field of the probe has been subtracted from the spectra.
Fig.\ref{fig2}(a) shows the mechanical-FMR spectra of the same disk at
four different frequencies. They substantially differ from the known
FMR signature of thin disks perpendicularly magnetized in the
saturated regime \cite{charbois02}.  These spectra usually consist of
an intense peak located at the highest field (lowest energy), the
uniform mode, followed by harmonics at lower field (higher energy),
which are standing spin-wave modes with an increasing order $m$ along
the radial direction. Such spectrum has been measured by cavity-FMR
technique on arrays of single Py layer disks with smaller thickness
($50$~nm) \cite{kakazei04}. This result has also been recently
confirmed by our mechanical-FMR setup on similar disks patterned out
of a $43.3$~nm single Py layer \cite{loubens06}.

Let us first concentrate on the FMR spectrum of Fig.\ref{fig2}(a)
measured at the highest frequency ($\omega/2\pi= 9.05$~GHz), when the
applied field is the largest, which insures that the magnetization is
the most homogeneous. The main peak occurs at $H_{u}\approx11.1$~kOe
(red square). Because it is the largest peak, this mode corresponds to
the largest volume of precession: the uniform (or $m=1$) mode. It is
followed by the first harmonic, $m=2$, at 10.4~kOe (open square),
whose size is consistent with its magnetostatic nature
\cite{loubens06}. In a 2D picture, the $m=2$ mode is characterized by
a nodal circle (locus where the magnetization does not precess) and an
outer cylindrical region precessing in opposition of phase with the
center. The striking new feature is the appearance of an additional
peak at $H_l=$11.4~kOe on the 9.05~GHz spectrum (blue triangle). This
new lowest energy mode has an amplitude (volume of precession), which
is about a factor of 5 smaller than the uniform mode. Repeating the
measurement on a second 1~$\mu$m disk confirms the reproducibility of
the spectral features (amplitude, position and linewidth) observed at
the highest frequency. The positions of the peaks follow a linear
frequency dependence in the measured range (see dashed lines in
Fig.\ref{fig2}(a)), a behavior consistent with the dynamics of a
saturated sample. (A slight departure from the fully saturated
configuration when $H_\text{ext} < 9.5$~kOe can be deduced from the
4.2~GHz spectrum, where the largest peaks occur below the red dashed
line). The linewidth of the uniform mode $\Delta H_u$ (half width at
half maximum) is only 28~Oe at 9.05~GHz.  This is among the
\emph{smallest} reported for Py.  Fig.\ref{fig2}(c) shows the
frequency dependence of the linewidth of the two lowest energy modes
for our micron-size disk. It varies linearly with frequency and the
slopes intercept with the origin. The zero intercept indicates that our
instrument resolves the \emph{intrinsic} broadening of the individual
mode. While there is usually a single peak at the resonance field of
the uniform precession \cite{loubens06}, our main mode (red square) is
split into at least two peaks. It becomes more obvious at lower
frequency, where the peaks are narrower and the signal to noise ratio
of the experiment improves. The same splitting of the largest peak is
also observed in the high frequency spectrum of the second 1~$\mu$m
disk. Such an observation points to the presence of significant
magnetic inhomogeneities inside the film of the disk sample (and not
to inhomogeneities in the lateral confinement, see below).

Quantitative prediction of the resonance field requires a detailed
knowledge of the magnetic properties of the Py layer. A control thin
film has been produced by placing a Si substrate inside the sputtering
chamber during the deposition process of the trilayer. Two basic FMR
studies have been performed on this Cu/NiFe/Cu extended film. The
first one is a 9.59~GHz cavity-FMR experiment where the resonance
field, $H_0$, is studied as a function of the orientation $\theta_H$
(angle with the film normal) of the static field $H_\text{ext}$.  The
second one is a stripline-FMR (thin film on top of a microwave
stripline), where $H_0$ is studied as a function of frequency at
$\theta_H=0$\textdegree{} (\textit{cf.}  Fig.\ref{fig1}(d)). The
analysis \cite{hurdequint02} of $H_0(\theta_H)$ gives a precise
determination of our Py layer $g$-factor, $g=2.134\pm0.003$
(\textit{i.e.}\nobreakspace a gyromagnetic ratio $\gamma=1.87~10^{7}
$~G$^{-1}$.s$^{-1}$) and $4\pi M_s=9801\pm1$~G for its saturation
magnetization at 298~K. These values correspond to a Ni$_{81}$Fe$_{19}$
alloy, although a decrease of $M_s$ should be assigned to the
adjunction of normal metal layers \cite{hurdequinticm}. 

This characterization can be used to understand the behavior observed
in the 1~$\mu$m disk.  We find that the slope $\gamma \partial
H_{u,l}/ \partial \omega = 1.03$ departs slightly from unity, what is
attributed to a small misalignment of about 2\textdegree{} between the
normal of the disk and $H_\text{ext}$ in the mechanical-FMR
experiment. This misalignment produces a shift down in field
($\approx -70$~G) of the spectra, which should be taken into account
when comparing it with the exact perpendicular configuration.
Following the same procedure as in Ref \cite{kakazei04}, we have
calculated the resonance fields of the magnetostatic modes $m=1,2,...$
at 9.05~GHz with a 2D analytical model, which neglects the
thickness dependence of the demagnetizing field. The value of $\gamma$ and
$M_s$ are inferred from the ones obtained on the extended film (no
fitting parameter), where $4 \pi M_s = 9.9$~kG is adjusted to the
value at the temperature of the mechanical experiment (280~K).
This model predicts that the mode $m=1$ should resonate at 11.07~kOe
and the mode $m=2$ at 10.45~kOe. Finer adjustment could be obtained by
fitting the pinning value of the magnetization at the disk periphery
\cite{kalinikos86}. Despite the simplifying assumptions and the
uncertainties of the precise experimental conditions (exact disk
diameter, misalignment of the field or strayfield of the probe), it
can account quite well for the main features observed in the low field
part of the spectrum of the disk.  But we also find that the 2D
assumption is unable to predict the lowest energy mode (triangle). It
requires a formalism which takes into account the variation of the
demagnetizing field in all three dimensions of the sample.

\begin{figure}
\includegraphics[width=10.0cm]{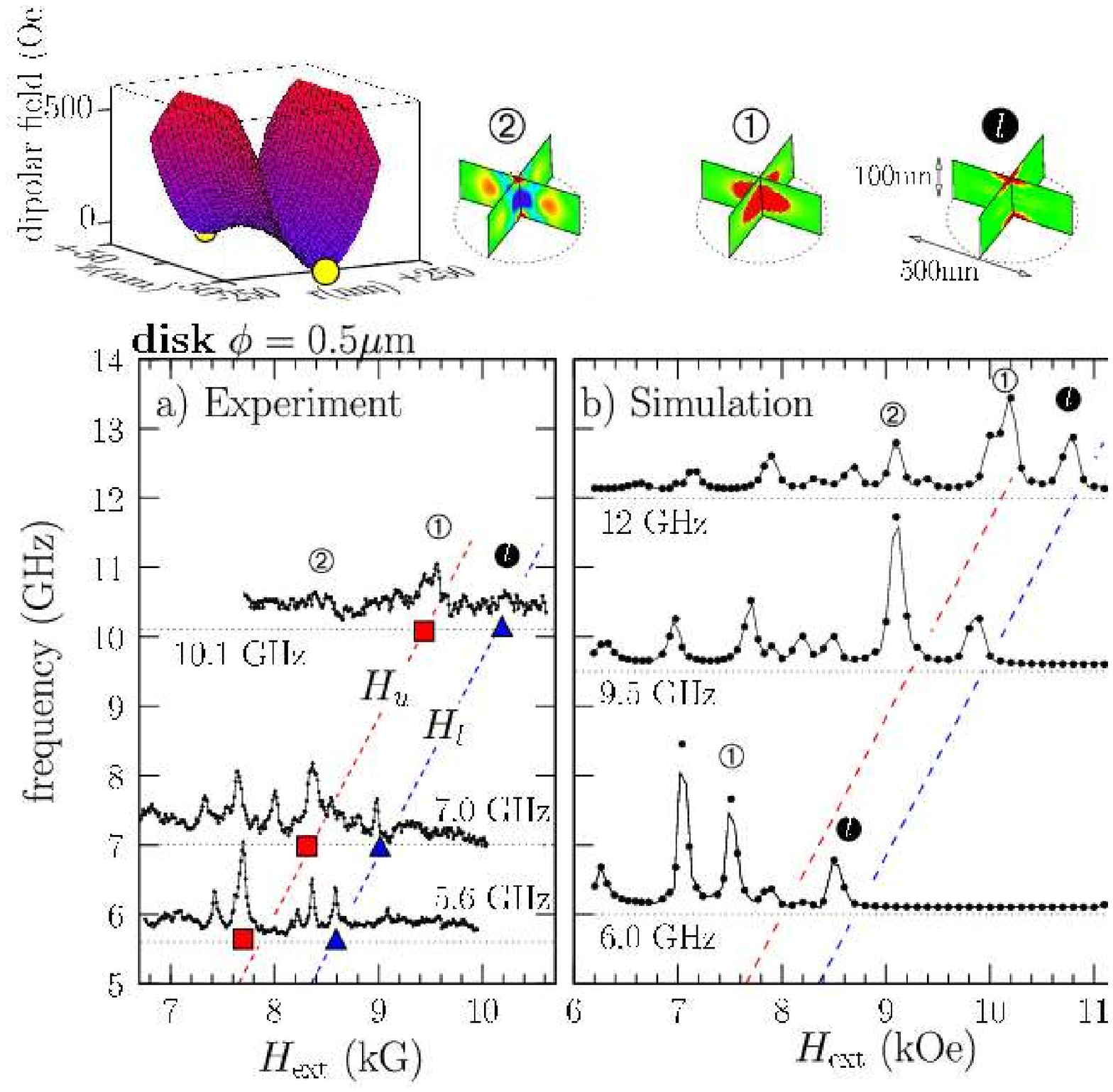}
\caption{(a) Mechanical-FMR spectra measured on the smallest disk
  ($\phi=0.5~\mu$m). (b) 3D simulation of the dynamical susceptibility
  of a single Py layer disk ($t=100$~nm, $\phi=0.5~\mu$m). The top
  shows the spatial dependence $(z,r)$ of the normal component of the
  demagnetizing field for the saturated configuration (shifted to zero
  at the disk center) and the transverse dynamics ($\chi''$) of the
  main modes in a color code.}
\label{fig3}
\end{figure}

To gain further insight, we have measured a smaller disk, 0.5~$\mu$m
in diameter. This aspect ratio departs further from the ellipsoidal
approximation used in the analytical model.  The result is shown in
Fig.\ref{fig3}(a). For the 10.1~GHz spectrum, the largest peak
(uniform mode) occurs here at 9.5~kOe (red square). There is again an
additional mode which resonates at lower energy: the peak at
$H_l=10.2~$~k0e on the same spectrum (triangle). Lower values of the
resonance fields are consistent with a decrease of the demagnetizing field
due to an increase of the aspect ratio ($t/ \phi$) and of the
quantization of the spin-wave modes along the diameter. The positions
of the peaks follow the same linear dependence on frequency above
7~GHz indicating that the onset of the fully saturated configuration
is around 8~kOe for this aspect ratio.

To describe the dynamics of the 0.5$~\mu$m disk, we have performed a
simulation of the dynamic susceptibility using the values of $M_s$,
$\gamma$ and damping (see below) measured experimentally.  $\chi''$ is
determined using two 3D codes developed by the authors \cite{boust04}.
The first one calculates a stable configuration of the magnetization
vector by solving the Landau-Lifshitz (LL) equation in the time
domain. The second one computes the full dynamic susceptibility tensor
from the linearization of the LL equation around the equilibrium
configuration. A mesh with a cubic cell of size 6~nm gives a discrete
representation of the sample. The role of the Cu interfaces is
neglected. We have plotted in Fig.\ref{fig3}(a) the amplitude of the
dynamical susceptibility at 12, 9.5 and 6 GHz. The cartography of
$\chi''$ is displayed using a color code for the first three modes.
The largest peak (the so-called ``uniform'' mode) corresponds to spins
that all precess in phase in the volume of the sample. A nodal circle
along the diameter appears in the next mode at lower field.  The
behavior in the median plan looks like the magnetostatic modes $m$
described in the 2D model, although we note the apparition of two
additional nodes across the sample thickness. The same visualization
of the lowest energy mode indicates that the mode $(l)$ corresponds to
a precession localized near the top and bottom surfaces of the
disk. The characteristic length scale of this precession along the
thickness is about 30~nm. The upper left part of Fig.\ref{fig3} shows
the calculated value of the $z$-component ($\| H_\text{ext}$) of the
demagnetizing field for the saturated configuration as a function of
the radial and thickness directions. The minima are located at the top
and bottom surfaces near the disk center (indicated by yellow
dots). Spins excited in the minima regions lead to a mode lower in
energy than the core precession \cite{jorzick02} because the gain in
demagnetizing energy exceeds the cost in exchange energy. Our
simulation computes precisely the competition between these two
interactions, but it neglects spin diffusion effects. The good
agreement with the data (no fitting parameters) indicates that these
are indeed the dominant effects. This picture is consistent with the
localized mode being absent of the spectrum of thinner samples, as the
ones used of Ref.  \cite{kakazei04} (it is also absent from the
simulation of a $t=43.3$~nm disk of same diameter).  One should note
that both the measurements and the simulations indicate a more complex
behavior as the unsaturated regime is approached (below 7~GHz). Part
of the observed features (split peaks) should be linked to the
imperfections identified in Fig.\ref{fig2}(a).

\begin{figure}
\includegraphics[width=10.0cm]{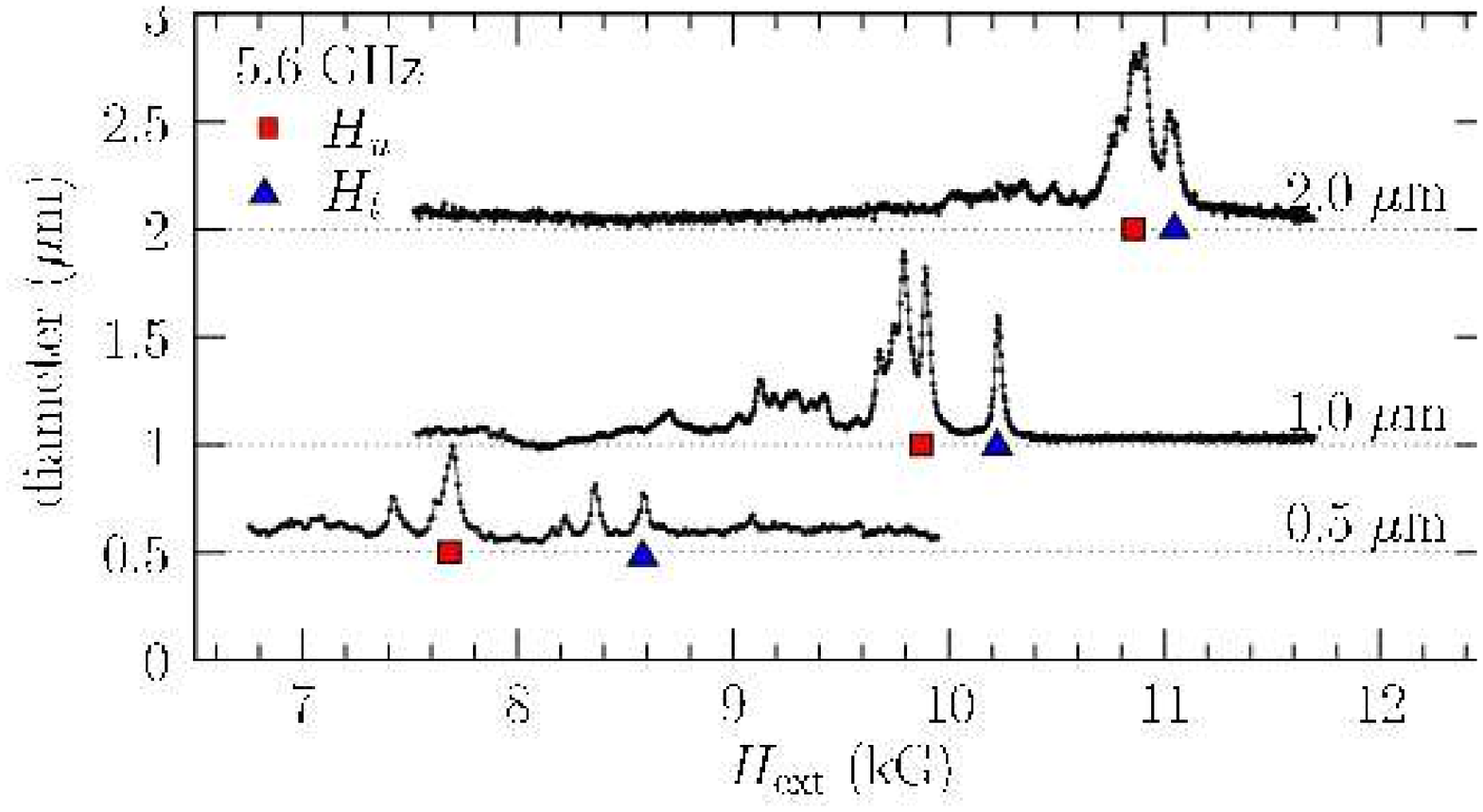}
\caption{Mechanical-FMR spectra of disks of different diameter at
  5.6~GHz.}
\label{fig4}
\end{figure}

Although this demagnetizing effect is clearly dependent on the thickness of
the Py layer, the amplitude and position of the localized mode vary
strongly with the lateral dimension of the disk. This is made evident
in Fig.\ref{fig4} where one compares the spectra obtained at the same
frequency ($5.6$~GHz) on three different diameters: 2, 1 and
0.5~$\mu$m.  The separation between the uniform and localized modes
increases when the diameter of the disk decreases, a behavior
consistent with a lift of degeneracy that is due to the confined
in-plane geometry. In summary, the detection of this localized mode
requires also to use small (submicron)-size disks.

Finally, we come back on the measured linewidths.  We observe in
Fig.\ref{fig2} and Fig.\ref{fig3} that the linewidth of the localized
mode, $\Delta H_l$, is smaller than $\Delta H_u$, the one of the
uniform mode.  Fig.\ref{fig2}(c) compares their frequency dependence
for the $\phi=1 ~\mu$m disk. The difference comes from the slopes
$\gamma \partial (\Delta H_{(u,l)}) / \partial \omega$, which yields
different damping coefficient $\alpha$, respectively $(1\pm0.1)
10^{-2}$ for the uniform mode and $(0.85\pm0.1)10^{-2}$ for the
localized mode.  Although the reason behind might be complicated, we
point out that lower dissipation is consistent with the fact that the
localized mode corresponds to the excitation of higher $k$-value
spin-waves \cite{loubens05}.

We have also investigated the cause of inhomogeneities in our disk.
Fig.\ref{fig2}(d) shows the frequency dependence of $\Delta H_0$, the
linewidth observed by stripline-FMR in the control film.  The
extrapolated value at zero frequency ($23\pm2$~G) shows the presence of
an inhomogeneous broadening in the multilayer structure
\cite{hurdequint02}, absent in the nanostructures. The detection of
the first volume spin-wave mode ($370$~Oe below the main mode) in the
stripline-FMR spectra (\textit{cf.}  Fig.\ref{fig1}(d)) indicates a
dynamical pinning behavior that can be described by the volume
inhomogeneity model \cite{wigen84} (a gradient of magnetization at the
Cu/NiFe interface).  This observation is consistent with a decrease of
$M_s$ by 25\% in a $\approx 10$~nm characteristic length scale at the
interface.  Although the study of the FMR spectrum of the control
film, deposited on a Si substrate, clearly reveals the presence of
such inhomogeneities, we indicate that the multilayer has been
deposited on the microstrip Au layer (7~nm roughness). Further studies
would be required to differentiate intrinsic contributions
(\textit{e.g.} spin diffusion into the normal metal layer
\cite{hurdequinticm}) to extrinsic ones, like interdiffusion of the
atomic species at the Cu/NiFe boundary or surface roughness of the
interfaces.

In conclusion, the above described mechanical-FMR technique has
enabled to perform measurement of spin-wave spectra in individual
submicron Cu/NiFe/Cu perpendicularly magnetized disks. Using 3D
simulations, we have shown that the lowest energy mode in a 100~nm
thick Py layer disk is localized in the region of the minimum
effective magnetic field (at the center of the Cu/Py
interface). Remarkably, the relaxation rate of this localized mode is
substantially smaller than the relaxation rate of the uniform mode. In
the case when the external bias field is fixed (\textit{e.g.} in the
case of spin-torque experiments) the localized mode has the lowest
relaxation rate both because it has the lowest damping coefficient and
because it is the lowest energy mode, and therefore it has the
smallest precession frequency.


We are greatly indebted to H. Hurdequint, A.N. Slavin, and A.-L.
Adenot for their help and support. This research was partially
supported by the ANR PNANO06-0235.

\vspace{-0.7cm}


\end{document}